# Compact microwave kinetic inductance nanowire galvanometer for cryogenic detectors at 4.2 K

S. Doerner, A. Kuzmin, K. Graf, S.Wuensch, I.Charaev, and M. Siegel

*Institut für Mikro- und Nanoelektronische Systeme, KIT, Hertzstr. 16, 76187, Karlsruhe, Germany*

We present a compact current sensor based on a superconducting microwave lumped-element resonator with a nanowire kinetic inductor, operating at 4.2 K. The sensor is suitable for multiplexed readout in GHz range of large-format arrays of cryogenic detectors. The device consists of a lumped-element resonant circuit, fabricated from a single 4-nm-thick superconducting layer of niobium nitride. Thus, the fabrication and operation is significantly simplified in comparison to state-of-the-art approaches. Because the resonant circuit is inductively coupled to the feed line the current to be measured can directly be injected without having the need of an impedance matching circuit, reducing the system complexity. With the proof-of-concept device we measured a current noise floor $\delta I_{min}$ of 10 pA/Hz$^{1/2}$ at 10 kHz. Furthermore, we demonstrate the ability of our sensor to amplify a pulsed response of a superconducting nanowire single-photon detector using a GHz-range carrier for effective frequency-division multiplexing.

Superconducting transition-edge sensors (TES) or superconducting nanowire single-photon detectors (SNSPD) are able to detect light on the single photon level over a wide spectral range. This makes them suitable for many research fields, e.g. the astronomy [1,2], particle physics [3] or material science [4,5]. Since their response after photon absorption is very weak, a sensitive preamplifier is essential.

In case of voltage biased TES very small current changes needs to be precisely amplified, which is realized in state-of-the-art systems using superconducting quantum interference devices (SQUIDs). SQUIDs allow the most sensitive readout possible. However, an SQUID based amplifier increases the system complexity and costs significantly. For next generation multi-pixel TES arrays also an array-scalable readout technique is required. This technique should reduce the number of cables between the low-temperature stage and the ambient-temperature back-end electronics, without compromising the sensitivity of the detectors. This is best realized using frequency-division multiplexing (FDM) approaches. So far, frequency-based multiplexing of TES arrays is realized by coupling the SQUIDs with additional resonant circuits [6,7] further increasing the complexity.

FDM is also very well suited for SNSPD arrays. However, to date there are very limited demonstrations of FDM in SNSPD arrays [8]. The most multiplexing approaches are based on single flux quantum (SFQ) logic schemes [9] or current splitting techniques [10] which as well increase the system complexity and reduce the filling factor of the sensing area.

An alternative approach to measure small currents or magnetic fields is based on nonlinear kinetic inductance $L_k$ in superconductors [11,12]. The nonlinearity could be considered quadratic on current $L_k/L_k(0) \approx 1+(I/I^*)^2$ for small current variations, where $I^*$ is the characteristic current, which depends on the critical current $I_C$ of the device [13]. Nowadays, this effect is being actively used for novel superconducting devices, e.g., tunable RF filters [14], low-noise wide-band parametric amplifiers [15]. Despite rather small current-driven changes of the kinetic inductance (≲ 15%) it is relatively easy to employ the effect. In this case it is beneficial to use superconductors with high kinetic inductance and geometries with a large fraction of kinetic inductance, for example nanowires made of ultra-thin NbN. The effect of the nonlinear $L_k$ in NbN nanowires was investigated previously in several works [16, 17]. Using superconducting resonators with high quality factors (high Q) the effect of the nonlinearity can be boosted proportionally to the Q factor. Potential applications of such devices are array-scalable magnetometers and current sensors, which could be an alternative to a SQUID-based readout schemes. Luomahaara *et. al.* demonstrated a kinetic inductance magnetometer based on a lumped-element NbN resonator, with a rather big inductor and a resonance frequency $f_{res}$ in the 100 MHz range. The magnetometer showed sensitivities comparable to SQUIDs [18]. There were also demonstrated GHz-range superconducting resonators with tunable kinetic nanoinductor using TiN [19]. The sensitivity was shown to be in the range of SQUIDs. But, because of the used material the operation temperature needs to be far below the boiling temperature of liquid helium. Also, their approach requires a large impedance matching circuit to feed the signal to be measured into the resonator, which for array applications may not be optimal. Another group demonstrated a tunable resonator based on NbTiN operated at 1.4 K [20]. However, to use their approach to multiplex arrays of detectors, a further impedance matching network is required.

As was shown in [21], nanowire inductors made of ultra-thin NbN are suitable for the use in compact GHz-range lumped-element resonators. The implementation of DC-biased NbN-nanowire inductors along with microwave high Q resonators, allows the fabrication of a sensitive current sensor which offers intrinsic FDM in array scalable applications. We call it the microwave kinetic inductance nanowire galvanometer (M-KING). Here, we present an experimental study of the device based on a microwave

Electronic mail: steffen.doerner@kit.edu or artem.kuzmin@kit.edu

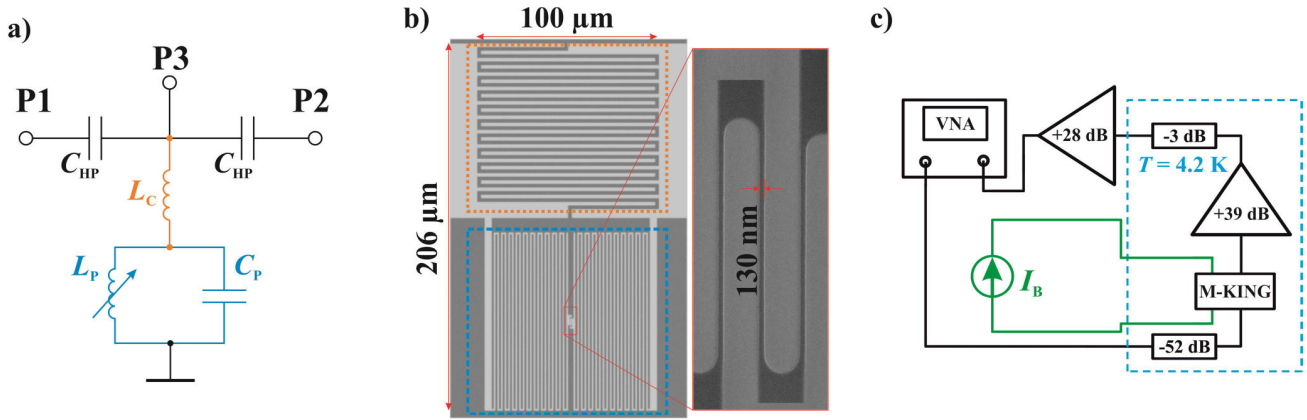

FIG. 1 (a) The equivalent circuit of the M-KING. The signal to be measured is feed through *P3* to GND. On its way it changes the kinetic inductance of $L_P$, which can be read out in a transmission measurement between $RF_{IN}$ and $RF_{OUT}$. (b) SEM Image of the M-KING device. The coupling inductor $L_C$ and the LC resonant circuit is highlighted. Enlarged on the right hand side the nanowire is shown, which adds the nonlinear behavior to the resonator. (c) Schematic of our setup used to measure the M-KING.

lumped-element superconducting resonator with a nanowire inductor working at 4.2 K. The M-KING is patterned in a single ultra-thin NbN layer, what makes the fabrication process significantly easier than for SQUIDs. The response of a DC-biased cryogenic detector can be directly injected into the resonator by a separate port and causes a change of its kinetic inductance, which is the dominant part of the total inductance ($L_p \approx L_{kin}$). A change of the injected current through the inductor consequently leads to a change of resonance frequency of the circuit. A small shift of the resonance frequency can be seen as a phase shift of a zero detuned microwave probe tone ($f_{tone} \approx f_{res}$) in the feed line. This approach allows for the FDM readout in GHz range of large arrays of cryogenic detectors.

### Design of M-KING

The design of the resonance circuit (FIG. 1a), which allows the direct injection of a current into the nonlinear inductor, differs from the previously demonstrated concepts [18, 19, 20]. The nonlinear inductance $L_P$ in parallel with the interdigital capacitor $C_P$ defines the resonance frequency of the resonator. The parallel circuit is coupled to a microwave feed line using a second lumped-element inductor $L_C$. This inductive coupling also represents a galvanic connection, which allows the injection of currents, in a wide range between DC up to several hundreds of MHz from the port P3 into the resonator. Thus, we do not need a further impedance matching circuit to couple the signal to be measured into the device. Everything is combined in the resonator design. The path in between P3 and the ground connection of the M-KING will hereinafter be referred to as DC path.

The layout was designed and simulated using the software *Sonnet EM* [22]. The nanowire $L_P$ is 19 μm long and 130 nm wide, which provides a kinetic inductance of 8.7 nH at zero bias. The capacitance $C_P$ is 237 fF and the coupling inductor $L_C$ has been set to 44.9 nH. The device is placed in the gap of a coplanar waveguide (CPW) shown in FIG 1(b). On the opposite gap of the CPW the tee-joint connection (P3) is placed to couple signals into the M-KING circuit which is only depicted in FIG 1(a). To separate the microwave path from the DC path, we used on the left and right side of M-KING two interdigital capacitors $C_{HP}$ which act as a galvanic decoupling. These capacitors eliminate leakage of the signal from P3 to other M-KING devices connected in parallel to the same feed line in a multi-pixel application. The simulated values of the resonance frequency $f_{res}$ and the loaded Q factor using typical parameters for NbN are $f_{res} \approx 3.8$ GHz and $Q_{\ell,sim} \approx 500$ respectively.

### Fabrication and measurements

The fabrication process starts with a reactive magnetron sputtering of 4-nm-thick NbN on single-side-polished sapphire substrate (R-plane) at a temperature of 850°C. The coplanar feed line and the M-KING circuit is patterned using electron-beam lithography on PMMA resist and subsequent $Ar^+$-ion etching. The fabricated proof-of-concept device is shown in FIG 1(b).

For measurements, the chip was mounted into a housing with three ports, two microwave ports to measure the transmittance of the feed line and one port for the injection of currents into the inductor. The scheme of the experimental setup is shown in FIG 1(c). The probing microwave signal from the Vector Network Analyzer (VNA) is applied to the input P1 of M-KING through a 52-dB cold attenuation line to suppress thermal noise. The transmitted microwave signal at port P2 is amplified by 38 dB using a cryogenic broadband high-electron-mobility transistor low-noise amplifier (HEMT LNA) with a noise temperature $T_{n1} \approx 6$ K. The total gain prior VNA is 75 dB and the equivalent noise temperature of the whole setup is $T_n \approx 26$ K. A third port P3 of the housing enables the injections of currents into the DC path of the M-KING. The housing with the chip was mounted into a dip-stick and cooled down in liquid helium to a temperature of 4.2 K. The measured critical current of the inductor $L_p$ is $I_{C,exp} \approx 33$ μA. During the measurements the bias current was swept up to 30 μA using a battery-driven current source. The change of the resonance frequency due to the applied bias current demonstrates the nonlinear behavior of the nanowire inductor (FIG. 2a). At a bias level of $0.9 I_{C,exp}$ we measured a resonance frequency shift $\approx 120$ MHz which is in good agreement with [20]. At zero bias the loaded quality factor is $Q_\ell \approx 600$, which is close to the simulated value.

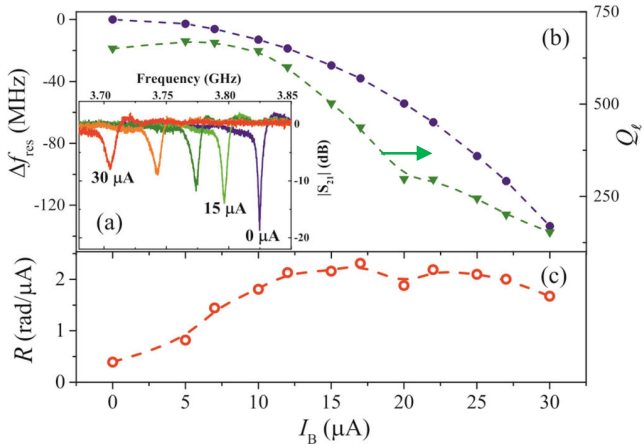

FIG. 2. Measured transmission over frequency of the M-KING circuit for different currents applied on the DC path at 4.2 K. The overall achieved shift of the resonance frequency is 119 MHz.

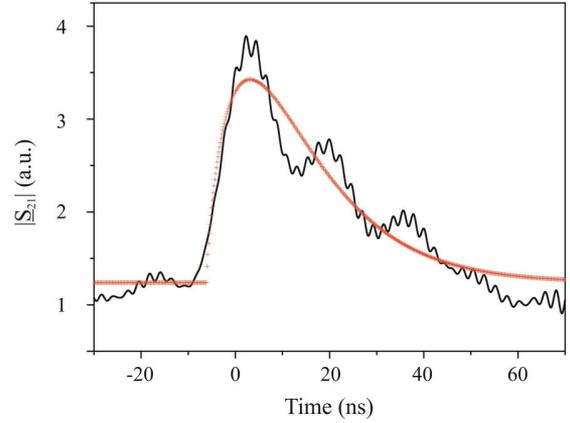

FIG 4 Measured transmittance on the feed line at resonance frequency of the M-KING. The changed transmittance is caused by a SNSPD pulse, which changes the resonance frequency and consequently increases the transmitted signal.

The small-signal phase responsivity $R_\theta = d\theta/dI$ was estimated for the zero-detuned probe using the measured current dependencies of the resonance frequency $f_{res}(I_b)$, the loaded quality factor $Q_\ell(I_b)$ shown in FIG 2(b) and the following expression:

$$R_{\theta 1} \approx (4Q_l/f_{res}) \times (df_{res}/dI) \quad (1)$$

Here, we used $d\theta \approx (4Q_l/f_{res}) \times df_{res}$ at resonance. The result is shown in FIG 2(c) and coincides with the value $R_{\theta 2} \approx 2\times 10^6$ rad/A, which was measured directly using a small DC-bias variation ($\Delta I \approx 100$ nA) and a fixed zero-detuned probe tone with the frequency $f_{probe} = f_{res}(15$ μA$)$. The spectrum of the current noise $\delta I$ of M-KING, shown

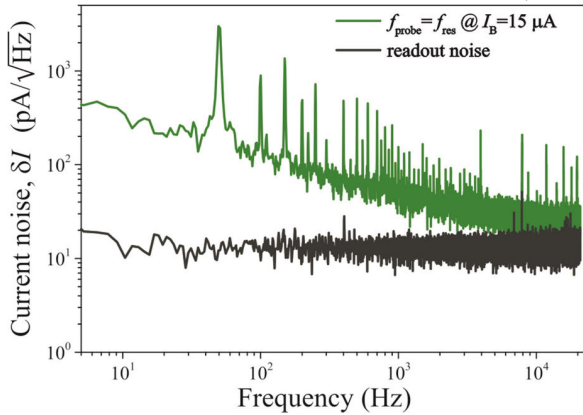

FIG 3 Measured spectrum of the current noise. The green curve was measured in resonance and the black curve was measured out of resonance. At a frequency above 10 kHz the current noise is about 10 pA/Hz$^{0.5}$

in FIG. 3 was recorded in the continuous-wave mode of the VNA as a time-trace of the phase signal at a bias level of 15 μA. A strong interference with the 50-Hz signal and its harmonics is clearly visible and could be suppressed by careful shielding and proper grounding. The readout noise floor was measured out of resonance and then normalized to the power levels in resonance. The noise equivalent current $\delta I$ reaches a minimum value of $\delta I_{min} \approx 10$ pA/Hz$^{1/2}$ at frequencies above 10 kHz, which are higher than the 1/f-noise corner frequency $f_C$. The achieved value is comparable with the sensitivity of previously demonstrated current sensing devices based on a nonlinear kinetic inductance [19] and is sufficient to readout signals from TES [23]. The excess 1/f-noise is due to a not optimized room-temperature DC-bias setup.

We also measured the response time of the M-KING directly using short pulses from a separate SNSPD in photon counting regime. The single-photon detector was mounted in a separate housing inside the dipstick and connected to the bias source on the one end and to the M-KING on the other. In case of a photon detection event, the SNSPD acts as a current switch. The rapid change of resistance causes a short current pulse, which is terminated by a capacitive coupled 50 Ohm resistor. Consequently, the current in the M-KING is lowered by almost 17μA which modulates the transmission of the M-KING's feed line. In FIG 4 such a modulated transmission response, recorded by a 32 GHz real-time oscilloscope is illustrated. After the superconducting state of the SNSPD is recovered, the M-KING also starts to recover its initial state. The required time for this change in frequency is determined by the resonance circuit ($\tau \approx Q/\pi f_{res}$).

### Discussion

In order to analyze the obtained experimental results, we derived the small-signal low-frequency phase responsivity $R_\theta = d\theta/dI$ and the noise for zero detuning, using the data of an electromagnetic simulation and the dependence of the kinetic inductance on current from Bardeen-Cooper-Schrieffer (BSC) or Ginsburg-Landau (GL) theories in the same way as in [16] and [17]. Assuming the inductance of the nanowire is dominated by the kinetic part and using (1) the responsivity can be expressed as $R_\theta \approx (2Q_l/L_k) \times (dL_k/dI)$. The theoretical current dependencies of the relative kinetic inductance $L_k(I)/L_k(0)$ and its normalized derivative $\varepsilon(I) = dL_k/dI \times I_C/L_k(0)$ are shown in FIG 5. The dependencies, derived from BSC and GL theories, are fitted to the experimentally measured data of M-KING. The phase responsivity now can be expressed as:

$$R_\theta \approx \frac{2Q_l(I) \times \varepsilon(I)}{I_C} \quad (2)$$

For the range of bias currents $I/I_C \approx 0.5$ the derivative $\varepsilon$ is approximately 0.1. Thus, the responsivity can be estimated as $R_\theta \approx 2 \times 10^6$ rad/A using the values of $Q_{l,\text{sim}} \approx 500$ and the measured $I_{C,\text{exp}} \approx 33$ μA, which is in very good agreement with the measurement. The estimate, derived from GL-theory is almost coinciding with BSC result. It is possible to express the critical current and Q factor of the device by the superconducting and geometrical properties of the nanowire for arbitrary case, like in [24] but expression (2) is more convenient to analyze in case of known typical parameters of the NbN nanowire (dependence $\varepsilon(I)$ is universal).

There are several noise sources in the M-KING: thermal noise of the microwave probe, readout phase noise, noise of the DC-bias source, a generation-recombination noise in the inductor and from two-level systems in the capacitor (TLS noise). We do not consider thermal noise of the microwave probe, since the 52-dB cold attenuator suppresses it significantly. TLS noise could be also ignored at $T=4.2$ K, as $f_{\text{res}} \ll k_B T/h$ ($k_B$ is the Boltzmann constant, $h$ – Plank's constant). The spectrum density of the phase noise due to microwave readout is $S_{\theta,\text{read}} = k_B T_n/P_{\text{out}}$, where $T_n$ is the effective noise temperature of the readout and $P_{\text{out}}$ is the microwave power on the output port P2 of M-KING. The spectrum density of the current noise due to readout is then:

$$\delta I_{\text{read}} = \sqrt{k_B T_n/P_{\text{out}}} \times R_\theta^{-1} \quad (3)$$

Using the responsivity of M-KING, the value of the microwave probe power at port P2 $P_{\text{out}} \approx 2$ pW and an effective noise temperature of the readout $T_n \approx 26$ K one can estimate the noise floor to be $\delta I_{\text{read}} \approx 8$ pA/Hz$^{1/2}$, which is in very good agreement with the experimentally measured value (FIG. 3). With an optimized back-end we can reduce the effective noise temperature of the system at least four times (down to $T_{n1} \approx 6$ K). It will give a factor of two lower noise floor $\delta I_{\text{read}} \approx 4$ pA/Hz$^{1/2}$.

As mentioned above, the used current source has a significant 1/f noise due to a not optimized layout. In our case the 1/f noise in the DC path is the dominant contribution at frequencies below 1 kHz (Fig. 3). This noise contribution can be significantly reduced using an improved bias source.

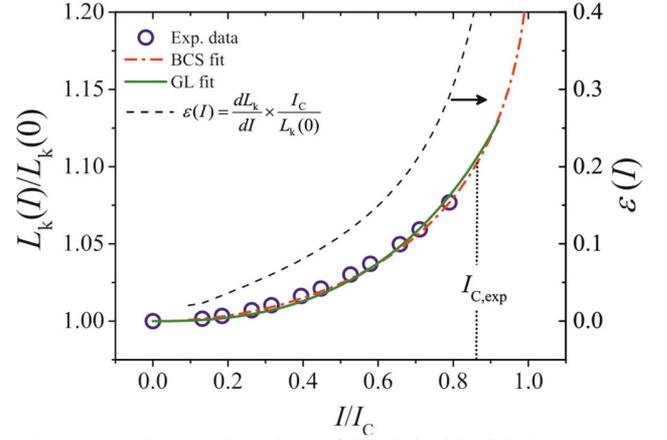

FIG 5. Measured current dependence of the relative kinetic inductance (open squares) fitted with BSC and GL theory. On the right axis shown the normalized derivative $\varepsilon(I)$, calculated from these fits.

At the finite operation temperature of the M-KING device a fluctuating number of thermally excited quasiparticles in the inductor can be considered as a random variation of the kinetic inductance. This will be translated into phase noise of M-KING according to $\delta\theta \approx (2Q_l/L_k)\delta L_k$. Small changes of the kinetic inductance can be estimated as $\delta L_k/L_k(0) \approx \delta N/N_s(0)$, where $N_s(0)$ is the total number of paired electrons in the inductor at $T = 0$ and $\delta N$ is the fluctuation in number of quasiparticles $\delta N \approx \sqrt{4N_{qp}(T)\tau_{qp}(T)}$ [25], where $N_{qp}(T)$, $\tau_{qp}(T)$ are mean number of quasiparticles and their life time correspondingly. Thus, the spectrum density of the phase noise due to quasiparticle generation-recombination processes is $S_{\theta,\text{g-r}} \approx 16Q_l^2 N_{qp}(T)\tau_{qp}(T)/N_s(0)^2$. The corresponding current noise could be estimated using (2) as:

$$\delta I_{\text{g-r}} \approx \frac{2I_C(T)}{N_s(0)\varepsilon(I)}\sqrt{N_{qp}(T)\tau_{qp}(T)} \quad (4)$$

For temperatures $T \ll T_c$ the product $N_{qp}(T)\tau_{qp}(T) = N_s(0)\tau_0/2\beta_0^3$ is temperature independent [26], where $\tau_0$ is a material characteristic time constant ($\tau_0 \sim 50$ ps [27] and $\beta_0 = \Delta_0/k_B T_c \approx 2.05$ for NbN). With this approximation and taking $\varepsilon \approx 0.1$, the volume of the inductor $V \approx 5 \times 10^{-21}$ m$^3$ and the density of the paired electrons in NbN $n_s(0) \sim 10^{29}$ m$^{-3}$ we can estimate $\delta I_{\text{g-r}} \sim 0.1$ pA/Hz$^{1/2}$, which is 2 orders of magnitude lower than the readout noise. It is worth to mention here that we neglect a contribution of the noise from superconducting fluctuations in form of vortex hoping, vortex-antivortex-pair unbinding or phase slips, which might dominate at bias currents >$0.9I_C$ (i.e. dark counts as in SNSPDs [28][29]).

## Conclusion

We designed, fabricated and characterized successfully a compact nanowire-based current sensor, M-KING, which demonstrated a current sensitivity of 10 pA/$\sqrt{\text{Hz}}$ at 4.2 K. The achieved sensitivity is close to what is required for the applications with TES bolometers. Moreover, the possibility to readout cryogenic detectors by converting their current response into microwave-resonance frequency shifts allows for FDM in GHz range. Along with a relatively easy fabrication process as well as the compact and flexible design it seems possible to build a large number of M-KINGs, which are connected to a common feed line, to readout large-format arrays. In case of SNSPD arrays, a multichannel M-KING opens a possibility to separate the SNSPD array from the multiplexer. Thus, the array can be optimized for high filling factors.

## Acknowledgment

This work was supported in part by the Karlsruhe School of Optics and Photonics (KSOP).